\documentclass[twocolumn,aps,showpacs,preprintnumbers]{revtex4}

\usepackage{amsmath,amsthm,amssymb}

\newtheorem{theorem}{Theorem}
\newtheorem{corollary}{Corollary}
\newcommand{\ket}[1]{\left|#1\right\rangle}
\newcommand{\bra}[1]{\left\langle#1\right|}

\begin{document}

\title{Quantum Contextuality in the Mermin-Peres Square: {A} Hidden
Variable Perspective\footnote{This work is an expanded version of a
paper to appear in \textit{Foundations of Probability and Physics-5},
G. Adenier, ed., AIP Conference Proceedings, 2009.}}

\author{Brian R. La Cour}
  \email{blacour@mail.utexas.edu}
  \affiliation{Applied Research Laboratories, 
               The University of Texas at Austin, 
               P.O. Box 8029, 
               Austin, TX 78713-8029}
\date{14 November 2008}

\begin{abstract}
The question of a hidden variable interpretation of quantum
contextuality in the Mermin-Peres square is considered.  The
Kochen-Specker theorem implies that quantum mechanics may be
interpreted as a contextual hidden variable theory.  It is shown that
such a hidden variable description can be viewed as either contextual
in the random variables mapping hidden states to observable outcomes
or in the probability measure on the hidden state space.  The latter
view suggests that this apparent contextuality may be interpreted as a
simple consequence of measurement disturbance, wherein the initial
hidden state is altered through interaction with the measuring device,
thereby giving rise to a possibly different final hidden variable
state from which the measurement outcome is obtained.  In light of
this observation, a less restrictive and, arguably, more reasonable
definition of noncontextuality is suggested.  To prove that such a
description is possible, an explicit and, in this sense, noncontextual
hidden variable model is constructed which reproduces all quantum
theoretic predictions for the Mermin-Peres square.  A critical
analysis of some recent and proposed experimental tests of
contextuality is also provided.  Although the discussion is restricted
to a four-dimensional Hilbert space, the approach and conclusions are
expected to generalize to any Hilbert space.
\end{abstract}

% 03.65.-w Quantum mechanics
% 03.65.Ta Foundations of quantum mechanics; measurement theory
% 03.65.Ud Entanglement and quantum nonlocality
% 03.67.-a Quantum information

% \keywords{}

\maketitle

%%%%%%%%%%%%%%%%%%%%%%%%%%%%%%%%%%%%%%%%%%%%%%%%%%%%%%%%%%%%%%%%%%%%%%%%%%%%%%%

\section{Introduction}

Contextuality is a property ascribed to quantum systems which appears to be at odds with a completely deterministic description.  However, due to its intrinsically counterfactual nature, the notion of contextuality can only be defined in terms of a hypothetical deterministic (i.e., hidden variable) description.  In the broadest sense, a measurement of an observable is said to be noncontextual if the outcome of the measurement does not depend upon which other compatible observables are measured subsequently, simultaneously, or previously.  The Kochen-Specker theorem purports to prove that no noncontextual hidden variable model exists which is consistent with quantum mechanics (for a Hilbert space of dimension three or greater).  Therefore, quantum mechanics is said to be contextual.

Spekkens \cite{Spekkens2005} has recently argued that such a definition of noncontextuality is overly restrictive since, as Bell observed much earlier, the particular outcome of a measurement may very well depend implicitly upon which other compatible observables are measured previously or simultaneously \cite{Bell1966}.  A better definition of a noncontextual measurement, then, would require only that the joint statistics of commuting observables be unchanged by the details of how they are measured.  For the present discussion, a noncontextual hidden variable model will be defined as one which associates a single random variable (i.e., measurable function) with each observable yet reproduces the correct joint statistics under a probability measure appropriate for the given experiment.

The Kochen-Specker theorem was first introduced by Bell \cite{Bell1966}, following his refutation of von Neumann's impossibility proof, and was itself based on the mathematical work of Gleason \cite{Gleason1957}.  It was independently proven by Kochen and Specker shortly afterwards \cite{Kochen&Specker1967} and became popular in philosophical circles.  The original theorem applied to Hilbert spaces of three dimensions, which may be viewed as describing the angular momentum of a spin-1 particle.  Simpler but more restrictive versions of the theorem in higher dimensions have since been published \cite{Greenberger1990,Mermin1990,Peres1991}.

Unlike Bell's inequality \cite{Bell1964}, the Kochen-Specker theorem is entirely nonstatistical --- in theory a single measurement suffices for empirical confirmation.  Thus, it is an example of an ``all-versus-nothing'' proof of the impossibility of hidden variables.  It was not until recently, however, that a  empirical test was proposed \cite{Simon2000}.  Subsequently, experiments using single and correlated photons \cite{Huang2003,Yang2005,Galvez2007} as well as neutron interferometry \cite{Hasegawa2006} have all shown results which are consistent with quantum theoretic predictions.  These experimental results appear to corroborate the theoretical prediction that quantum mechanics is inescapably contextual.

The aim of this paper is to demonstrate that such a conclusion is unwarranted and that, in fact, quantum theory is perfectly consistent with a deterministic, noncontextual theory.  To this end, it will be argued that the standard proofs of the Kochen-Specker theorem become invalid when one allows for the possible dependence of the \emph{post-measurement} hidden variable probability distribution on the particular set of mutually commensurate observables chosen for measurement.  Furthermore, it will be shown how this dependence may arise naturally through the process of measurement and attendant interactions with the measuring devices.  To illustrate the concept and demonstrate that such a scheme is possible, an explicit noncontextual hidden variable model is constructed and studied for observations on a system prepared in an entangled state.  Although the discussion is restricted to a four-dimensional Hilbert space, the approach and conclusions are expected to generalize to any Hilbert space.

Early attempts to explain quantum phenomena by appealing to measurement interactions failed in their inability to provide physically reasonable models, consistent with known classical laws and relevant interaction time scales \cite{Jammer1966}.  This implies either new physics or, as Bell has suggested, lack of imagination \cite{Bell1987}.  A careful analysis of the original Kochen-Specker theorem by Belinfante \cite{Belinfante1973} showed that a contradiction could be avoided if one assumed that the measurement outcome depends explicitly on the particular set of commuting operators (of which the given observable is a member).  Such a possibility was rejected as explicitly contextual; here it is argued that such a dependence may arise via measurement interaction.  Meyer, Kent, and Clifton \cite{Meyer1999,Kent1999,Clifton2000} have argued that inescapably finite measurement precision effectively nullifies the Kochen-Specker result, a conclusion which has been criticized by Appleby and others (q.v., \cite{Appleby2002} and references within).  Since then, several experimental tests of contextuality have been performed, indicating that finite precision is not the issue.  More recently, Leifer and Spekkens have formulated a connection between contextuality and certain pre- and post-selection (PPS) paradoxes \cite{Leifer&Spekkens2005}.  While they observe that the concept of measurement disturbance can resolve PPS paradoxes, they fall short of applying this same reasoning to contextuality itself.

The organization of the paper is as follows.  In Sec.\ \ref{sec:PS} a particular representation of the four-dimensional Hilbert space is introduced.  The Kochen-Specker theorem is considered in the context of nine composite Pauli spin operators on this space, and the proof for this case is shown to be invalid if one relaxes certain, arguably unwarranted, assumptions.  The reasons for this conclusion are further elaborated upon in Sec.\ \ref{sec:HVP}, where a hidden variable interpretation is offered.  Sec.\ \ref{sec:IM} provides an illustrative noncontextual model for measurements that are either sequential or simultaneous.  A critical analysis of some recent and proposed experimental tests follows in Sec.\ \ref{sec:DE}, where the concept of operator decomposability is introduced.  A summary and final conclusions are given in Sec.\ \ref{sec:SC}.

%%%%%%%%%%%%%%%%%%%%%%%%%%%%%%%%%%%%%%%%%%%%%%%%%%%%%%%%%%%%%%%%%%%%%%%%%%%%%%%

\section{Problem Statement}
\label{sec:PS}

A general, four-dimensional Hilbert space may be mapped to a notional composite system of two spin-1/2 particles.  For the single-particle component subspace, any self-adjoint operator may be written as a linear combination of the Pauli spin operators, $\hat{\sigma}_x$, $\hat{\sigma}_y$, $\hat{\sigma}_z$, and the identity, $\hat{1}$.  The discussion that follows will be cast in terms of this representation.

Let us begin by considering the example of the Mermin-Peres ``magic square'' \cite{Mermin1990,Peres1991}, which consists of nine operators arranged as follows:
\begin{equation*}
\begin{array}{lll}
\hat{\sigma}_x \otimes \hat{1} \quad & \hat{1} \otimes \hat{\sigma}_x \quad & \hat{\sigma}_x \otimes \hat{\sigma}_x \\
\\
\hat{1} \otimes \hat{\sigma}_y \quad & \hat{\sigma}_y \otimes \hat{1} \quad & \hat{\sigma}_y \otimes \hat{\sigma}_y \\
\\
\hat{\sigma}_x \otimes \hat{\sigma}_y \quad & \hat{\sigma}_y \otimes \hat{\sigma}_x \quad & \hat{\sigma}_z \otimes \hat{\sigma}_z
\end{array}
\end{equation*}

Let $\hat{V}_{ij}$ denote the operator in row $i$, column $j$.  From the properties of the Pauli spin operators it is readily verified that the three operators in each row are mutually commuting, as are those in each column.  Furthermore, it can be shown that each operator \emph{anticommutes} with the four operators not in its row or column.  Finally, we observe that the product of the three operators in each row, as well as those in the first two columns, is $+\hat{1}\otimes\hat{1}$.  The product of the operators in Column 3, by contrast, is $-\hat{1}\otimes\hat{1}$.

For a noncontextual model of the Mermin-Peres magic square, we seek a set of hidden variables, $\Omega$, and a collection of nine functions $V_{ij}: \Omega \to \mathbb{R}$ (for $i,j = 1,2,3$) such that the outcome of measuring operator $\hat{V}_{ij}$ is $V_{ij}(\omega)$, where $\omega \in \Omega$.  Suppose $\Omega$ and all nine $V_{ij}$ are given.  The aforementioned operator relations for the product of each row and column \emph{suggest} a similar relation in the hidden variable model.

For $i=1,2,3$, let us define $R_i$ to be the set of all $\omega \in \Omega$ such that $V_{i1}(\omega), V_{i2}(\omega), V_{i3}(\omega) \in \{-1,+1\}$ and
\begin{equation}
\label{eqn:Ri}
V_{i1}(\omega) V_{i2}(\omega) V_{i3}(\omega) = +1.
\end{equation}
Similarly, for $j=1,2$, let $C_j$ be the set of all $\omega \in \Omega$ such that $V_{1j}(\omega), V_{2j}(\omega), V_{3j}(\omega) \in \{-1,+1\}$ and
\begin{equation}
\label{eqn:Cj}
V_{1j}(\omega) V_{2j}(\omega) V_{3j}(\omega) = +1.
\end{equation}
Finally, let $C_3 \subseteq \Omega$ be such that, for all $\omega \in C_3$, $V_{13}(\omega), V_{23}(\omega), V_{33}(\omega) \in \{-1,+1\}$ but, by contrast, 
\begin{equation}
\label{eqn:C3}
V_{13}(\omega) V_{23}(\omega) V_{33}(\omega) = -1.
\end{equation}

Now, let us suppose that there exists at least one point, $\omega$, that is common to all six row/column sets.  From Eqn.\ (\ref{eqn:Ri}) it follows that
\begin{equation}
\prod_{i=1}^{3} V_{i1}(\omega) V_{i2}(\omega) V_{i3}(\omega) = (+1)(+1)(+1) = +1.
\end{equation}
Furthermore, from Eqns.\ (\ref{eqn:Cj}) and (\ref{eqn:C3}) it follows that
\begin{equation}
\prod_{j=1}^{3} V_{1j}(\omega) V_{2j}(\omega) V_{3j}(\omega) = (+1)(+1)(-1) = -1.
\end{equation}
But
\begin{equation}
\prod_{i=1}^{3} V_{i1}(\omega) V_{i2}(\omega) V_{i3}(\omega) = \prod_{j=1}^{3} V_{1j}(\omega) V_{2j}(\omega) V_{2j}(\omega),
\end{equation}
so we arrive at a contradiction and conclude that
\begin{equation}
\label{eqn:NCI}
(R_1 \cap R_2 \cap R_3) \cap (C_1 \cap C_2 \cap C_3) = \varnothing.
\end{equation}

Thus far, we have not incorporated quantum theory other than to suggest the form for $R_i$ and $C_j$.  The standard proof of the Kochen-Specker theorem assumes that the functional relations held by the operators imply that
\begin{equation}
\label{eqn:QMA}
R_1 = R_2 = R_3 = C_1 = C_2 = C_3 = \Omega.
\end{equation}
In other words, that Eqns. (\ref{eqn:Ri})--(\ref{eqn:C3}) hold for \emph{all} $\omega \in \Omega$.  If this is so, then Eqn.\ (\ref{eqn:NCI}) implies that $\Omega = \varnothing$ and no (non-vacuous) noncontextual model is possible.

The proof is almost trivial, but it relies on one key assumption: the validity of Eqn.\ (\ref{eqn:QMA}).  Greenberger \textit{et al.}\ \cite{Greenberger1990} have noted that such equalities are overly restrictive, as the statistical nature of quantum mechanics requires only that each set have unit probability measure.  This is true, but a more subtle observation, which has largely gone unnoticed, is that the relevant probability measure may in fact be different for each of the six sets.  How this is possible, and what it implies, are the subject of the following sections.

%%%%%%%%%%%%%%%%%%%%%%%%%%%%%%%%%%%%%%%%%%%%%%%%%%%%%%%%%%%%%%%%%%%%%%%%%%%%%%%

\section{Hidden Variable Perspective}
\label{sec:HVP}

In this section we consider the magic square problem from the point of view of hidden variable theory.  It will be shown that the desired functions $V_{ij}$ may be constructed, consistent with all quantum predictions, provided one allows that the probability measure may vary with the measurement context.  As will be discussed in Sec.\ \ref{ssec:UC} and illustrated in Sec.\ \ref{sec:IM}, this apparent contextual dependence can arise as a natural consequence of the measurement process through the interaction between the measurement apparatus and the system under investigation.  In this view, then, $V_{ij}(\omega)$ is \emph{not} the value of the observable prior to measurement but, rather, its value \emph{after} having interacted with the measuring device.

%------------------------------------------------------------------------------

\subsection{Probability Theoretic Description}

It is well known that for a set of mutually commuting operators, $\{\hat{A}_1, \hat{A}_2, \hat{A}_3, \ldots\}$, and a quantum state $\hat{\rho}$ (which may be pure or mixed) in a given Hilbert space there exists a (non-unique) probability space $(\Omega, \mathcal{F}, Q)$ and set of real-valued, $\mathcal{F}$-measurable functions (i.e., random variables) $\{A_1, A_2, A_3, \ldots\}$ on $\Omega$ such that their joint distribution under $Q$ reproduces the joint distribution of the corresponding operators under $\hat{\rho}$ \cite{vonNeumann,Gudder}.

The aforementioned probability space is determined, in part, by the quantum state and set of operators, but it is neither uniquely nor completely specified by them.  For example, the probability measure, $Q$, may be defined on, but assigned zero probability to, values of the random variables that fall outside the spectrum of the operators.  The nature of the sample space, $\Omega$, is, of course, quite arbitrary.  

A common, often implicit, assumption made in proofs of quantum contextuality is the so-called \emph{Functional Composition Principle} \cite{Redhead1987}.  Suppose the aforementioned operators satisfy the functional relation
\begin{equation}
\hat{f}(\hat{A}_1, \hat{A}_2, \hat{A}_3, \ldots) = \hat{0}
\end{equation}
for some function $\hat{f}$ on the operator space, where $\hat{0}$ is the null operator.  Applying the same functional form of $\hat{f}$ to a function $f$ on the domain of reals, it can be shown that the corresponding random variables satisfy a similar relation,
\begin{equation}
\label{eqn:BSA}
f(A_1(\omega), A_2(\omega), A_3(\omega), \ldots) = 0,
\end{equation}
for $Q$-almost every $\omega \in \Omega$ (i.e., over a set whose probability measure under $Q$ is one).  Unfortunately, this latter qualifier is often ignored in discussions of hidden variables, and so Eqn.\ (\ref{eqn:BSA}) is often taken to hold for \emph{all} $\omega \in \Omega$.  The distinction is not merely pedantic.  Indeed, it is this assumption which gives rise to Eqn.\ (\ref{eqn:QMA}) and the conclusion that $\Omega = \varnothing$.  More generally, regions of $\Omega$ may have unit probability for one set of mutually commuting operators yet zero probability for another.  This observation is fundamental to understanding the Kochen-Specker theorem.

In summary, for a given set of mutually commuting operators and a particular quantum state, one can always construct a corresponding hidden variable model, though perhaps not a very interesting one.  It follows that for two such sets, two separate hidden variable models (i.e., probability spaces and associated random variables) can be constructed.  The following subsections describe how these separate constructs can be combined into one and how this, in turn, can be used to construct a set of noncontextual random variables.

%------------------------------------------------------------------------------

\subsection{Contextual Random Variables}

In the previous subsection we observed that, for a given quantum state, each set of mutually commuting operators gives rise to a corresponding probability space and set of associated random variables.  Each of the six row/column sets constitutes such a set.  We shall denote the corresponding probability space and set of random variables by $(\Omega_n, \mathcal{F}_n, Q_n)$ and $\{A_{1|n}, A_{2|n}, A_{3|n}\}$, with $n = 1, 2, 3$ for Rows 1, 2, 3 and $n = 4, 5, 6$ for Columns 1, 2, 3, respectively.

Suppose we wish to combine the six measurable spaces, $(\Omega_1, \mathcal{F}_1), \ldots, (\Omega_6, \mathcal{F}_6)$, into a single, common measurable space $(\Omega, \mathcal{F})$.  Mathematically, it is easy to construct such a space in terms of a product space by defining
\begin{subequations}
\begin{align}
\Omega &:= \Omega_1 \times \cdots \times \Omega_6, \\
\mathcal{F} &:= \mathcal{F}_1 \otimes \cdots \otimes \mathcal{F}_6,
\end{align}
\end{subequations}
where $\mathcal{F}$ is the product $\sigma$-algebra \cite{Dudley}.

On this space we may define the projection map $\pi_n: \Omega \to \Omega_n$ (for $n = 1, \ldots, 6$) such that, for any $\omega = (\omega_1, \ldots, \omega_6) \in \Omega$,
\begin{equation}
\pi_n(\omega) := \omega_n.
\end{equation}
With this definition, the random variables on each $\Omega_n$ may be extended to $\Omega$ by simply composing them with the corresponding projection map.  Thus, we define
\begin{equation}
A'_{m|n} := A_{m|n} \circ \pi_n
\end{equation}
for variable $m \in \{1, 2, 3\}$ and context $n \in \{1, \ldots, 6\}$.

Finally, a common probability measure may be defined as the product measure $P = Q_1 \times \ldots \times Q_6$.  Such a distribution reproduces all quantum theoretic predictions for a given row/column set but may give nonsensical results otherwise.  For example, consider the joint distribution of $A'_{1|1}$ and $A'_{1|4}$.  Since $P$ is a product measure, the two random variables are explicitly independent, yet the corresponding operator, $\hat{\sigma}_x\otimes\hat{1}$, is the same for both.  The random variables are thus explicitly contextual, even though the probability measure is noncontextual.  In the following section we will consider a different representation which takes into account the possible overlap between such sets and, thereby, allows one to define noncontextual random variables.

%-----------------------------------------------------------------------------

\subsection{Contextual Probability Measures}

The random variables $A'_{1|1}$ and $A'_{1|4}$ correspond to the same operator, $\hat{\sigma}_x \otimes \hat{1}$, but in different contexts.  The former is measured in conjunction with the other two in its row, $\hat{1} \otimes \hat{\sigma}_x$ and $\hat{\sigma}_x \otimes \hat{\sigma}_x$, while the latter is measured in conjunction with the other two in its column, $\hat{1} \otimes \hat{\sigma}_y$ and $\hat{\sigma}_x \otimes \hat{\sigma}_y$.  To avoid such contextual dependencies, we may define a random variable $V_{11}: \Omega \to \mathbb{R}$ such that
\begin{equation}
V_{11}(\omega) := 
\begin{cases}
A'_{1|1}(\omega), & \text{if $A'_{1|1}(\omega) = A'_{1|4}(\omega)$}, \\
0, & \text{otherwise}.
\end{cases}
\end{equation}
The alternative value of 0 is arbitrary, so long as it occurs with zero probability.  By definition, $V_{11}$ is noncontextual yet always yields the same value that either $A'_{1|1}$ or $A'_{1|4}$ would obtain.  We may define all nine $V_{ij}$ (for $i,j = 1,2,3$) in a similar manner.

Finally, we may specify probability measures $P_1, \ldots, P_6$ on $(\Omega, \mathcal{F})$ based on the known probability measures $Q_1, \ldots, Q_6$ on $(\Omega_1, \mathcal{F}), \ldots, (\Omega_6, \mathcal{F}_6)$, respectively.  To account for the fact that, say, $A'_{1|1}$ and $A'_{1|4}$ refer to the same operator, we may define $P_1$ such that
\begin{equation}
\begin{split}
P_1[A'_{1|1}&=a_{1|1}, \ldots, A'_{3|6}=a_{3|6}] := \\
&\delta(a_{1|1}, a_{1|4}) \delta(a_{2|1}, a_{1|5}) \delta(a_{3|1}, a_{1|6}) \\
&\times Q_1[A_{1|1}=a_{1|1}, A_{2|1}=a_{2|1}, A_{3|1}=a_{3|1}] \\
&\times Q_2[A_{1|2}=a_{1|2}, A_{2|2}=a_{2|2}, A_{3|2}=a_{3|2}] \\
&\times Q_3[A_{1|3}=a_{1|3}, A_{2|3}=a_{2|3}, A_{3|3}=a_{3|3}] \\
&\times Q_4[A_{2|4}=a_{2|4}, A_{3|4}=a_{3|4}] \\
&\times Q_5[A_{2|5}=a_{2|5}, A_{3|5}=a_{3|5}] \\
&\times Q_6[A_{2|6}=a_{2|6}, A_{3|6}=a_{3|6}],
\end{split}
\label{eqn:CPP}
\end{equation}
where $\delta(x,y) = 1$, if $x=y$, and $0$ otherwise.  The remaining five probability measures may be defined similarly.

Each $P_n$ refers to one of the six mutually commuting sets and, hence, to a different random experiment.  In this sense, the probability measures are contextual.  Nevertheless, using the above definition for $P_1$, and a similar one for $P_4$, we find that the two corresponding marginals are independent of context.  For example, consider
\begin{equation}
\begin{split}
P_1[V_{11}=\pm1] &= P_1[A'_{1|1}=\pm1, A'_{1|4}=\pm1] \\
&= Q_1[A_{1|1}=\pm1] \\
&= Q_4[A_{1|4}=\pm1] \\
&= P_4[A'_{1|1}=\pm1, A'_{1|4}=\pm1] \\
&= P_4[V_{11}=\pm1],
\end{split}
\end{equation}
where $Q_1[A_{1|1}=\pm1] = Q_4[A_{1|4}=\pm1]$ is a standard quantum theoretic result.  If, however, one considers only measurements of operators in the first row, then
\begin{multline}
% \begin{equation}
P_1[V_{11}=v_{11}, \, V_{12}=v_{12}, \, V_{13}=v_{13}] \\
= Q_1[A_{1|1}=v_{11}, \, A_{2|1}=v_{12}, \, A_{3|1}=v_{13}],
% \end{equation}
\end{multline}
as expected.  Finally,  mixed context probabilities such as
\begin{equation}
P_1[V_{11}=v_{11}, V_{12}=v_{12}, V_{21}=v_{21}]
\end{equation}
are also well defined, though perhaps meaningless.

As a consequence of these results, for a set such as $R_1$, we have
\begin{equation}
\begin{split}
P_1[R_1] &= P_1[V_{11} \, V_{12} \, V_{13} = +1] \\
&= Q_1[A_{1|1} \, A_{2|1} \, A_{3|1} = +1] = 1,
\end{split}
\end{equation}
where the final equality derives from quantum theory.  A similar result holds for the other five sets, and we conclude that
\begin{equation}
\begin{split}
P_1[R_1] = P_2[R_2] = P_3[R_3] &= 1, \\
P_4[C_1] = P_5[C_2] = P_6[C_3] &= 1.
\end{split}
\label{eqn:CUP}
\end{equation}
Thus, even though each set occurs almost surely in the context of its corresponding experiment, there is no point common to all six sets, as was shown earlier.

To summarize, it has been shown that it is possible to construct a set of nine random variables, $V_{ij}$, such that their common domain, $\Omega$, is nonempty and the subsets $R_1, \ldots, C_3$ have probability 1, albeit with respect to different measures.  This set, together with the corresponding probability measures, reproduces all statistical predictions of quantum theory for the magic square problem.  The resulting hidden variable model is noncontextual in the random variables; however, it is effectively contextual in the probability measures.

Whether the random variables or probability measures are viewed as contextual, the question remains how this apparent contextual dependence arises.  This question will be addressed in the following section, where we will find that contextuality, as such, can arise naturally through the process of measurement.

%------------------------------------------------------------------------------

\subsection{Understanding Contextuality}
\label{ssec:UC}

Given an $\omega_0 \in \Omega$, we know that it will \emph{not} be contained in at least one of the six row/column sets $R_1, \ldots, C_3$.  If it happens to be the case that $\omega_0 \in R_1$, then $V_{11}(\omega_0) \, V_{12}(\omega_0) \, V_{13}(\omega_0) = +1$, as one might expect.  If, however, it happens to be the case that $\omega_0 \not\in C_1$, then we find, perhaps surprisingly, that $V_{11}(\omega_0) \, V_{21}(\omega_0) \, V_{31}(\omega_0) \neq +1$.  Since $\omega_0$ was arbitrary, the question arises why this is never observed.

One possible answer lies in a taking a closer look at the measurement process itself.  Measuring an observable such as $\hat{V}_{11} = \hat{\sigma}_x\otimes\hat{1}$ requires a particular apparatus designed to interact with the system under interrogation.  This process need not be benign.  Suppose $\omega_0 \in \Omega$ describes the initial \emph{microstate} of the system.  (Here the term ``system'' may refer not only to the specific object of inquiry but also to the measuring device, surrounding environment, etc.)  Interaction with the measuring apparatus may cause it to change its microstate from $\omega_0$ to some $\varphi_{11}(\omega_0) \in \Omega$.  Let us call this function the \emph{measurement interaction map} (MIM).  An observation then maps this microstate to some \emph{macrostate} $g_{11}(\varphi_{11}(\omega_0)) \in \mathbb{R}$.  If we consider an \emph{ensemble} of initial microstates described by the probability measure $P_0$, then the ensemble after interaction becomes $P_0 \circ \varphi_{11}^{-1}$.  The distribution for the macrostate is then $P_0 \circ \varphi_{11}^{-1}\circ g_{11}^{-1}$.

Although described as an artificial discrete map, the transformation of the system microstate should more properly be viewed as a continuous interaction process for which $\omega_0$ and $\varphi_{11}(\omega_0)$, say, represent asymptotic (i.e., interaction-free) initial and final values.  This would imply that a characteristic time scale exists over which the interactions must take place in order to agree with quantum mechanical predictions.  In this view, $g_{11}(\varphi_{11}(\omega_0))$ represents the long-time asymptotic macrostate.  The existence of such a time scale, and its observed magnitude, would place important constraints on any proposed hidden variable model.

The situation is quite different in classical statistical mechanics, where the process of extracting a macrostate from the system is often ignored or irrelevant.  Thus, $g_{11}(\omega_0)$ may be the true macrostate of the system prior to measurement, but we cannot observe it directly.  Though we may \emph{hypothesize} its existence, we may only \emph{access} it via measurement.  The process of measurement, however, results in our measuring $g_{11}(\varphi_{11}(\omega_0))$ which, depending upon the nature of $\varphi_{11}$, may not be the same as $g_{11}(\omega_0)$.  (In the logical positivist philosophical tradition, one may go further and assert that $g_{11}(\omega_0)$, having no operational definition, is simply meaningless.)

If a subsequent measurement of, say, $\hat{V}_{12} = \hat{1}\otimes\hat{\sigma}_x$ is made, a similar process unfolds.  The microstate $\varphi_{11}(\omega_0)$ is now transformed into $\varphi_{12}(\varphi_{11}(\omega_0))$, and the observed macrostate is $g_{12}(\varphi_{12}(\varphi_{11}(\omega_0)))$.  The ensemble is transformed in a like manner from $P_0 \circ \varphi_{11}^{-1}$ to $P_0 \circ \varphi_{11}^{-1} \circ \varphi_{12}^{-1}$, and the joint distribution of the two measurements is therefore $P_0 \circ (g_{11} \circ \varphi_{11}, \; g_{12} \circ \varphi_{12} \circ \varphi_{11})^{-1}$.  Had we chosen to measure $\hat{V}_{21} = \hat{1}\otimes\hat{\sigma}_y$ instead of $\hat{1} \otimes \hat{\sigma}_x$, the observed macrostate would have been $g_{21}(\varphi_{21}(\varphi_{11}(\omega_0)))$, and the final ensemble would have been $P_0 \circ \varphi_{11}^{-1} \circ \varphi_{21}^{-1}$.

Of course, simultaneous measurements may also be possible, in which case the MIMs $\varphi_{11}$, $\varphi_{12}$, and $\varphi_{13}$, say, will be replaced by a single MIM, $\Phi_1$.  Similarly, $\varphi_{11}$, $\varphi_{21}$, and $\varphi_{31}$, will be replaced by a single $\Phi_4$.  In this case, the ensemble following a measurement of Row 1 will be $P_0 \circ \Phi_1^{-1}$, while that of Column 1 will be $P_0 \circ \Phi_4^{-1}$.  Letting $G_{ij}$ denote the macroscopic map corresponding to the operator $\hat{V}_{ij}$, a measurement of, say, Row 1 results in the values $G_{1j}(\Phi_1(\omega_0))$ for $j = 1,2,3$, while a measurement of, say, Column 1 results in the values $G_{i1}(\Phi_4(\omega_0))$ for $i = 1,2,3$.  It is an academic matter whether one considers the random variable $V_{11} = G_{11}$ as noncontextual, with the contextual probability measures $P_1 = P_0 \circ \Phi_1^{-1}$ and $P_4 = P_0 \circ \Phi_4^{-1}$, or whether one considers $A'_{1|1} = G_{11} \circ \Phi_1$ and $A'_{1|4} = G_{11} \circ \Phi_4$ as contextual, with $P = P_0$ now noncontextual.

If such interactions do indeed exist, their effect must be consistent with the statistical predictions of quantum mechanics.  It is desirable that they also satisfy our various intuitive notions of physical realism.  For example, if a measurement of $\hat{V}_{11}$ is followed by a time-like separated measurement of either $\hat{V}_{12}$ or $\hat{V}_{21}$, we expect the outcome of the first measurement to be independent of which observable is chosen for the second measurement.  Furthermore, the outcome of measuring $\hat{V}_{11}$ should not depend upon whether $\hat{V}_{12}$ is measured before $\hat{V}_{13}$ or $\hat{V}_{13}$ is measured before $\hat{V}_{12}$.  Indeed, this should be true even if a measurement of $\hat{V}_{11}$ is followed by a measurement of an incompatible observable, such as $\hat{V}_{22}$.  Since, according the above description, the outcome of the first measurement is always $g_{11}(\varphi_{11}(\omega_0))$, all these conditions are clearly satisfied.

Now suppose $\hat{V}_{12}$ is measured first, followed by a time-like measurement of $\hat{V}_{11}$.  Should we demand that the outcome of the latter, namely $g_{11}(\varphi_{11}(\varphi_{12}(\omega_0)))$, be identical to the outcome that would have resulted if $\hat{V}_{11}$ were measured first, namely $g_{11}(\varphi_{11}(\omega_0))$?  This is certainly possible, but it is unreasonable and unwarranted to demand it.  Although the joint distributions must be the same, i.e.,
\begin{multline}
% \begin{equation}
P_0 \circ (g_{11} \circ \varphi_{11}, \; g_{12} \circ \varphi_{12} \circ \varphi_{11})^{-1} \\
= P_0 \circ (g_{11} \circ \varphi_{11} \circ \varphi_{12}, \; g_{12} \circ \varphi_{12})^{-1},
% \end{equation}
\end{multline}
it is not necessary that $g_{11} \circ \varphi_{11} = g_{11} \circ \varphi_{11} \circ \varphi_{12}$, nor $g_{12} \circ \varphi_{12} = g_{12} \circ \varphi_{12} \circ \varphi_{11}$, in order for this to be true \cite{Spekkens2005}.  Thus, as long as the correct quantum statistics are reproduced, this constraint is also satisfied.

Finally, if a measurement of $\hat{V}_{11}$ is repeated, even after measurements of other compatible observables have been made, then quantum theory predicts (and observation dictates) that the same outcome must be obtained.  Thus, for example, we require that $g_{11} \circ \varphi_{11} = g_{11} \circ \varphi_{11} \circ \varphi_{12} \circ \varphi_{11}$ $P_0$-almost everywhere.  This, and similar relations, do place important constraints on a noncontextual hidden variable model and reflect, in part, the von Neumann postulate regarding wavefunction collapse.

Of course, if the measurements are simultaneous and not co-located, or merely space-like separated, local realism imposes more severe constraints.  If $\hat{\sigma}_x \otimes \hat{1}$ and $\hat{1} \otimes \hat{\sigma}_x$, say, represent spin measurements on two distant spin-1/2 particles, then we certainly would expect that relations such as $g_{11} \circ \varphi_{11} \circ \varphi_{12} = g_{11} \circ \varphi_{11}$ hold exactly and not just in their distributions.  This, and similar relations, place severe constraints on the choice of MIMs consistent with a local hidden variable theory.  A nonlocal, noncontextual theory of space-like separated measurements is, however, still possible, as the following section illustrates.

%%%%%%%%%%%%%%%%%%%%%%%%%%%%%%%%%%%%%%%%%%%%%%%%%%%%%%%%%%%%%%%%%%%%%%%%%%%%%%%

\section{Illustrative Model}
\label{sec:IM}

This section provides an illustrative, albeit contrived, example of a set of deterministic microscopic interaction maps and macroscopic functions which produce the apparent contextual behavior of the quantum magic square.  First, sequential measurements are considered.  (Here, the term ``sequential'' is understood to mean either time-like separated or light-like separated and not co-located.)  Using this, a model for simultaneous (i.e., space-like separated or light-like and co-located) measurements is then constructed.  The section ends with several examples of measurements on an entangled system \footnote{This model has been implemented in Matlab\textsuperscript{\textregistered} by the author and is freely available at the following URL: \textsf{https://webspace.utexas.edu/blacour/www/}.}.

%------------------------------------------------------------------------------

\subsection{Sequential Measurements}

Let $\Omega = [0,1]^{\infty}$ be the infinite-dimensional unit hypercube, let $\mathcal{F}$ be the set of Borel subsets of $\Omega$, and let $P_0 = \mu$ be Lebesgue measure on $\mathcal{F}$.  Thus, $(\Omega, \mathcal{F}, \mu)$ is a probability space \cite{Kolmogorov1933}.   A element of $\Omega$ represents the hidden variable ``microstate'' and will be denoted $u = [u_1, u_2, \ldots]$, where $u_i \in [0,1]$ for each $i \in \mathbb{N}$.  As will be explained, each component of $u$ encodes either information about a past measurement or information needed to predict a future measurement.

For this example, each measurement will transform one of the coordinates to the finite set $\{0/27, \ldots, 26/27\}$ in a manner to be described below.  Writing $27u_1$ in the base-three notation $(d_2 d_1 d_0)_3$, the most significant digit, $d_2$, represents the outcome of a particular measurement.  (An outcome of $-1$ corresponds to $d_2 = 0$, while and outcome of $+1$ corresponds to $d_2 = 2$.)  The middle and least significant digits encode the row and column, minus 1, of the observable measured.  Thus, a value of $u_1 = (012)_3/27 = 5/27$ signifies an initial measurement of $\hat{V}_{23}$ was performed with an outcome of $-1$. 

With this in mind, let us define the set
\begin{multline}
% \begin{equation}
M_n := \{ u \in [0,1]^{\infty} :\; 27u_k\:\mathrm{mod}\:1 = 0 \; \forall \; k \le n \;\mathrm{and}\; \\
\mbox{   }27u_{k}\:\mathrm{mod}\:1 \neq 0 \; \forall \; k > n \} \; .
% \end{equation}
\end{multline}
The initial hidden variable will, with $\mu$-probability 1, lie in $M_0$.  After the first measurement, it will transform to a value in $M_1$, etc.  The actual transformation of $u$ is effected as follows.

Given an initial quantum state $\hat{\rho}$, measurement of the operator $\hat{V}_{ij}$ will follow a cumulative distribution function (CDF) $F_{ij}: \mathbb{R} \to [0,1]$ determined from the Born rule.  Thus, $F_{ij}(-1)$ is the probability of obtaining $-1$ upon an initial measurement of $\hat{V}_{ij}$, while $F_{ij}(+1) = 1$.  This CDF will be used to construct the first MIM.  Now, given the outcome of the first measurement, and the operator measured, one may use the projection postulate to define the conditional CDF $F_{ij}(\cdot|u_1)$ for a subsequent measurement of another, possibly different, operator.  (Recall that $u_1$ contains the relevant information regarding the previous measurement.)  Continuing in this manner, the conditional CDF for the $n$th measurement will be $F_{ij}(\cdot|u_1, \ldots, u_{n-1})$.  With this in mind, the nine MIMs are defined as follows.
\begin{multline}
% \begin{equation}
\varphi_{ij}(u) := \\
\begin{cases}
[\nu_{ij}(\chi_{ij}(u)), u_2, \ldots] & \text{if $u \in M_0$} \\
[u_1, \ldots, u_{n-1}, \nu_{ij}(\chi_{ij}(u)), u_{n+1}, \ldots] & \text{if $u \in M_n$} \\
[u_1, u_2, \ldots] & \text{otherwise}
\end{cases}
% \end{equation}
\end{multline}
where the measurement outcome is given by
\begin{multline}
% \begin{equation}
\chi_{ij}(u) := \\
\begin{cases}
\mathrm{inf}\{ x : F_{ij}(x) > u_1 \} & \text{if $u \in M_0$} \\
\mathrm{inf}\{ x : F_{ij}(x|u_1, \ldots, u_{n-1}) > u_n\} & \text{if $u \in M_n$}
\end{cases}
% \end{equation}
\end{multline}
and the measurement history is encoded via
\begin{equation}
\nu_{ij}(x) := 
\begin{cases}
[3(i-1) + j-1]/27 & \text{if $x = -1$} \\
[18 + 3(i-1) + j-1]/27 & \text{if $x = +1$} \; .
\end{cases}
\end{equation}
Finally, the outcome of the measurement is given by the macroscopic map $g: [0,1]^{\infty} \to \{-1, 0, +1\}$, defined by
\begin{equation}
\label{eqn:littleg}
g(u) := 
\begin{cases}
\mathrm{sign}(2u_n-1) & \text{if $u \in M_n$} \\
0 & \text{otherwise},
\end{cases}
\end{equation}
where $n$ is the largest positive integer such that $u \in M_n$.  This model will be referred to as the Sequential Measurements model.

Thus, given $u \in M_0$, an initial measurement of, say, $\hat{V}_{11}$ results in the outcome $g(\varphi_{11}(u))$, while a subsequent measurement of, say, $\hat{V}_{12}$ will yield $g(\varphi_{12}(\varphi_{11}(u)))$.  Now, had $\hat{V}_{21}$ been chosen for measurement instead of $\hat{V}_{12}$, the result would have been $g(\varphi_{21}(\varphi_{11}(u)))$.  Furthermore, had the incompatible operator $\hat{V}_{22}$ been chosen instead of either $\hat{V}_{12}$ or $\hat{V}_{21}$, the result would be given by $g(\varphi_{22}(\varphi_{11}(u)))$.  In all three cases, the joint distribution of the two measurements is manifestly identical to that predicted by quantum theory.  Furthermore, it is clear that the outcome of the first measurement, namely $g(\varphi_{11}(u))$, is independent of which observable is measured subsequently.  More generally, we have the following theorems, which are proven in the Appendix.

\begin{theorem}
The Sequential Measurements model reproduces all statistical predictions of quantum mechanics for all possible sequential measurements of operators from the magic square.
\label{thm:Ordering}
\end{theorem}

\begin{corollary}
Under the Sequential Measurements model, if a measurement is repeated, and no intervening incompatible measurements are performed, then the outcome will $\mu$-almost surely remain unchanged.
\label{cor:Persistence}
\end{corollary}

%------------------------------------------------------------------------------

\subsection{Simultaneous Measurements}

The previous section considered a model of sequential measurements, wherein the outcome of each measurement depends explicitly upon the past measurement history.  Here we consider extending the Sequential Measurements model to measurements which are simultaneous (in some inertial frame).  This extension will be referred to as the Simultaneous Measurements model.  The significance of such measurements is that a single MIM is applied to the initial microstate, after which the appropriate macroscopic maps are applied to obtain the measurement outcomes.

One obvious way to construct such a MIM is to compose a set of MIMs for a particular sequence of time-like separated measurements.  This is possible since, by Corollary \ref{cor:Persistence}, the same outcome of a measurement for a particular observable will always be obtained upon repeating that measurement.  Thus, for example,
\begin{equation}
g \circ \varphi_{11} = g \circ \varphi_{11} \circ \varphi_{12} \circ \varphi_{11}
\end{equation}
$\mu$-almost everywhere.  If, however, the order of the measurements is reversed, such an equality may no longer hold; i.e., over a set of $\mu$-probability greater than zero
\begin{equation}
g \circ \varphi_{11} \neq g \circ \varphi_{11} \circ \varphi_{12} \; .
\end{equation}
So the choice of a particular ordering is important and may correspond, for example, to a particular experimental arrangement or set of devices.  (Note that each choice of ordering defines a different, albeit equivalent, Simultaneous Measurements model.)

This suggests that we may define $G_{ij} := g \circ \varphi_{ij}$ and $\Phi_n$ by, say, the following:
\begin{subequations}
\begin{align}
\Phi_{1} &:= \varphi_{11} \circ \varphi_{12} \circ \varphi_{13} \quad (\text{Row 1}) \\
\Phi_{2} &:= \varphi_{21} \circ \varphi_{22} \circ \varphi_{23} \quad (\text{Row 2}) \\
\Phi_{3} &:= \varphi_{31} \circ \varphi_{32} \circ \varphi_{33} \quad (\text{Row 3}) \\
\Phi_{4} &:= \varphi_{11} \circ \varphi_{21} \circ \varphi_{31} \quad (\text{Column 1}) \\
\Phi_{5} &:= \varphi_{12} \circ \varphi_{22} \circ \varphi_{32} \quad (\text{Column 2}) \\
\Phi_{6} &:= \varphi_{13} \circ \varphi_{23} \circ \varphi_{33} \quad (\text{Column 3})
\end{align}
\end{subequations}
The ordered triple of outcomes in a simultaneous or space-like separated measurement of, say, Row 1 is then
\begin{equation*}
[G_{11}, \, G_{12}, \, G_{13}] \circ \Phi_1.
\end{equation*}
In general, $G_{11} \circ \Phi_1 \neq G_{11} \circ \Phi_4$ and $G_{11} \circ \Phi_1 \neq g \circ \varphi_{11}$, even though the statistical distribution is the same for all three random variables.  As discussed previously, this is not a violation of noncontextuality but merely a reflection of the possible dependence of a \emph{particular} outcome on the experimental procedure.  In a realistic (i.e., non-contrived) hidden variable theory, the probability space $(\Omega, \mathcal{F}, P_0)$ may well be entirely different for each such procedure.  Note also that, in contrast to the explicitly local Sequential Measurements model, the form of $\Phi_n$ assumed in the Simultaneous Measurements model incorporates an explicit nonlocality, unless the measurement interactions are light-like separated and co-located.

With these definitions we may now make associations with the random variables and probability measures discussed in Sec.\ \ref{sec:HVP}.  Thus, the noncontextual random variable $V_{ij}$ may be identified with the macroscopic map $G_{ij}$, while the contextual probability measure $P_n$ may be identified with $\mu \circ \Phi_n^{-1}$.  Similarly, the noncontextual probability measure, $P$, may be identified with $\mu$, while the contextual random variable $A'_{m|n}$ may be identified with either $G_{n,m} \circ \Phi_n$ (for $n = 1,2,3$) or $G_{m,n-3} \circ \Phi_n$ (for $n = 4,5,6$).

%%%%%%%%%%%%%%%%%%%%%%%%%%%%%%%%%%%%%%%%%%%%%%%%%%%%%%%%%%%%%%%%%%%%%%%%%%%%%%

\subsection{Some Examples with an Entangled State}
\label{sec:EE}

Suppose a physical system is prepared in an entangled state given in the $z$-basis by
\begin{equation}
\hat{\rho} = \frac{1}{2} \begin{bmatrix}
0 & 0 & 0 & 0 \\
0 & +1 & -1 & 0 \\
0 & -1 & +1 & 0 \\
0 & 0 & 0 & 0
\end{bmatrix}.
\end{equation}
For definiteness, suppose the hidden variable state is
\begin{equation}
\label{eqn:IM}
u = [0.76, \; 0.51, \; 0.02, \; 0.82, \; u_5, \; u_6, \ldots] \in M_0,
\end{equation}
where the first four numbers are exact, not approximate, and the remaining $u_5, u_6, \ldots$ are left unspecified.  (Recall that $\mu(M_0) = 1$, so taking $u$ to be in $M_0$ is quite reasonable.)

%-----------------------------------------------------------------------------

\subsubsection{Experiment 1: Sequential Measurements of Row 3}

In this experiment, we measure first $\hat{V}_{33}$, then $\hat{V}_{32}$, then finally $\hat{V}_{31}$.  These correspond to the three operators $\hat{\sigma}_z \otimes \hat{\sigma}_z$, $\hat{\sigma}_y \otimes \hat{\sigma}_x$, and $\hat{\sigma}_x \otimes \hat{\sigma}_y$, respectively, in Row 3.  The initial microstate $u$, prior to measuring $\hat{\sigma}_z \otimes \hat{\sigma}_z$, is given by Eqn.\ (\ref{eqn:IM}), and the initial, unobserved macrostate corresponding to this operator is $0$, according to Eqn.\ (\ref{eqn:littleg}).

After the first measurement, the microstate is altered to the value
\begin{equation}
u' := \varphi_{33}(u) = [\tfrac{8}{27}, 0.51, 0.02, 0.82, u_5, u_6, \ldots],
\end{equation}
and the observed macrostate (i.e., the measurement outcome) is $g(u') = -1$.  A subsequent measurement of $\hat{\sigma}_y \otimes \hat{\sigma}_x$ now alters the microstate from $u'$ to $u''$, where
\begin{equation}
u'' := \varphi_{32}(u') = [\tfrac{8}{27}, \tfrac{25}{27}, 0.02, 0.82, u_5, u_6, \ldots],
\end{equation}
and the observed macrostate is $g(u'') = +1$.  Finally, we measure $\hat{\sigma}_x \otimes \hat{\sigma}_y$ and obtain
\begin{equation}
u''' := \varphi_{31}(u'') = [\tfrac{8}{27}, \tfrac{25}{27}, \tfrac{6}{27}, 0.82, u_5, u_6, \ldots],
\end{equation}
with an outcome of $g(u''') = -1$.  As expected, the product of the three outcomes is $+1$.  By Theorem \ref{thm:Ordering}, this will be true for $\mu$-almost any choice of $u$.

%-----------------------------------------------------------------------------

\subsubsection{Experiment 2: Sequential Measurements of Column 3}

In this experiment, we measure first $\hat{V}_{33}$, then $\hat{V}_{23}$, then finally $\hat{V}_{13}$.  These correspond to the three operators $\hat{\sigma}_z \otimes \hat{\sigma}_z$, $\hat{\sigma}_y \otimes \hat{\sigma}_y$, and $\hat{\sigma}_x \otimes \hat{\sigma}_x$, respectively, in Column 3.

Using the same initial microstate given by Eqn.\ (\ref{eqn:IM}), the post-measurement microstate, $u'$, and observed macrostate, $g(u') = -1$, are of course the same as in Experiment 1.  Following a measurement of $\hat{\sigma}_y \otimes \hat{\sigma}_y$, however, the microstate becomes
\begin{equation}
u'' := \varphi_{23}(u') = [\tfrac{8}{27}, \tfrac{5}{27}, 0.02, 0.82, u_5, u_6, \ldots],
\end{equation}
and the measurement outcome is $g(u'') = -1$.  A final measurement of $\hat{\sigma}_x \otimes \hat{\sigma}_x$ now yields
\begin{equation}
u''' := \varphi_{13}(u'') = [\tfrac{8}{27}, \tfrac{5}{27}, \tfrac{2}{27}, 0.82, u_5, u_6, \ldots],
\end{equation}
and the measurement outcome is $g(u''') = -1$.  The product of the outcomes is $-1$, as expected.

%------------------------------------------------------------------------------

\subsubsection{Experiment 3: Simultaneous Measurement of Row 3}

In this experiment, we measure the operators in Row 3 simultaneously using $\Phi_3$.  The post-measurement microstate is now
\begin{equation}
u' := \Phi_3(u) = [\tfrac{8}{27}, \tfrac{25}{27}, \tfrac{6}{27}, 0.82, u_5, u_6, \ldots],
\end{equation}
and the observed values are
\begin{equation}
[G_{31}(u'), \; G_{32}(u'), \; G_{33}(u')] = [-1, \; +1, \; -1].
\end{equation}
As it turns out, these results are exactly the same as those in Experiment 1.

Had we instead performed a simultaneous measurement of Column 3, we would have obtained
\begin{equation}
v' := \Phi_6(u) = [\tfrac{8}{27}, \tfrac{5}{27}, \tfrac{2}{27}, 0.82, u_5, u_6, \ldots],
\end{equation}
and
\begin{equation}
[G_{13}(v'), \; G_{23}(v'), \; G_{33}(v')] = [-1, \; -1, \; -1].
\end{equation}

Note that, in the case of $v'$, the \emph{unobserved} macrostates of Row 3 are different from those of $u'$.  Specifically,
\begin{equation}
[G_{31}(v'), \; G _{32}(v'), \; G_{33}(v')] = [+1, \; +1, \; -1],
\end{equation}
and their product, as it turns out, does \emph{not} equal $+1$.  Similarly, the unobserved macrostates of Column 3 in the case of $u'$ are
\begin{equation}
[G_{13}(u'), \; G _{23}(u'), \; G_{33}(u')] = [+1, \; +1, \; -1],
\end{equation}
which differ from those of $v'$ but, as it turns out, still have a product of $-1$.  Thus, while $u$ is contained in neither $R_3$ nor $C_3$, we find that $u' \in R_3$ and $v' \in C_3$.  Again, by Theorem \ref{thm:Ordering}, this will be true for $\mu$-almost every $u$.  Note that there is also a nonzero probability (with respect to $\mu$) that $u$ will be such that both $u'$ and $v'$ lie in $R_3 \cap C_3$.

%------------------------------------------------------------------------------

\subsubsection{Experiment 4: Alternative Measurement of Row 3}

In Experiment 3 we found that the results of a simultaneous measurement of either Row 3 or Column 3 were precisely those found in the sequential measurements of Experiments 1 and 2, respectively.  This need not be so.  Consider a variation of Experiment 3 wherein we measure the operators in Row 3 simultaneously using a different device, one for which the MIM is
\begin{equation}
\Phi'_{3} := \varphi_{33} \circ \varphi_{32} \circ \varphi_{31}.
\end{equation}
In this case, the post-measurement microstate is
\begin{equation}
u' := \Phi'_3(u) = [\tfrac{24}{27}, \tfrac{7}{27}, \tfrac{8}{27}, 0.82, u_5, u_6, \ldots]
\end{equation}
and the measurement outcomes are
\begin{equation}
[G_{31}(u'), \; G_{32}(u'), \; G_{33}(u')] = [+1, \; -1, \; -1].
\end{equation}
While it remains true, as it must, that the product of the outcomes is $+1$, the particular values obtained for this device happen to be different.

%%%%%%%%%%%%%%%%%%%%%%%%%%%%%%%%%%%%%%%%%%%%%%%%%%%%%%%%%%%%%%%%%%%%%%%%%%%%%%

\section{Operator Decomposability and Some Recent Experiments}
\label{sec:DE}

A frequent assumption made in discussions of contextuality is that, in a noncontextual theory, the outcome of a measurement on the product of two commuting operators, commonly written $v[\hat{A} \hat{B}]$, is equal to the product, $v[\hat{A}] v[\hat{B}]$, of the outcomes that would have been obtained had either of the two operators been measured individually \cite{Cabello1998}.  This section critically examines the basis for this assumptions.

Consider the operators $\hat{\sigma}_x \otimes \hat{1}$ and $\hat{1} \otimes \hat{\sigma}_y$ from the magic square.  For a given $\omega \in \Omega$ we may make the following associations:
\begin{subequations}
\begin{align}
v[\hat{\sigma}_x \otimes \hat{1}] &= V_{11}(\omega) \\
v[\hat{1} \otimes \hat{\sigma}_y] &= V_{21}(\omega) \\
v[\hat{\sigma}_x \otimes \hat{\sigma}_y] &= V_{31}(\omega) \; .
\end{align}
\end{subequations}
Now, the aforementioned assumption is that $V_{31}(\omega) = V_{11}(\omega) V_{21}(\omega)$.  By definition, this equality holds if and only if $\omega \in C_1$, and this, in turn, will almost always hold whenever a measurement of Column 1 is performed.  (Recall that, as per the discussion of Sec.\ \ref{ssec:UC}, $\omega$ is interpreted as the post-measurement microstate.)  If, however, a measurement of, say, Row 1 is performed, it may well be that $\omega \not\in C_1$, in which case the assumption of equality may be false.

In this section we will consider conditions under which one may legitimately decompose the noncontextual random variables into a product of constituent random variables.  We will then turn to consider the implications of this decomposability property for some recent experimental tests of quantum contextuality.

%------------------------------------------------------------------------------

\subsection{Decomposability}

Each of the nine operators in the magic square may be written in terms of the four basic operators $\hat{\sigma}_x\otimes\hat{1}$, $\hat{1}\otimes\hat{\sigma}_x$, $\hat{\sigma}_y\otimes\hat{1}$, and $\hat{1}\otimes\hat{\sigma}_y$. This raises the question of whether it is possible to write each of the nine noncontextual random variables in terms of the four basic random variables $V_{11}$, $V_{12}$, $V_{22}$, $V_{21}$, which we shall denote here by $X_1$, $X_2$, $Y_1$, $Y_2$, respectively.

From the definitions of the six row/column sets, we note the following:
\begin{subequations}
\begin{align}
\omega \in R_1 \;\Rightarrow\; V_{13}(\omega) &= X_1(\omega) X_2(\omega) \\
\omega \in R_2 \;\Rightarrow\; V_{23}(\omega) &= Y_1(\omega) Y_2(\omega) \\
\omega \in R_3 \;\Rightarrow\; V_{33}(\omega) &= V_{31}(\omega) V_{32}(\omega) \\
\omega \in C_1 \;\Rightarrow\; V_{31}(\omega) &= X_1(\omega) Y_2(\omega) \\
\omega \in C_2 \;\Rightarrow\; V_{32}(\omega) &= X_2(\omega) Y_1(\omega) \\
\omega \in C_3 \;\Rightarrow\; V_{33}(\omega) &= -V_{13}(\omega) V_{23}(\omega) \; .
\end{align}
\end{subequations}
Since $\omega$ is not contained in at least one of these six sets, we have at most five equations to define the five remaining unknowns.  If $\omega$ is contained in only four or fewer sets, then a full decomposition may not be possible.  If, however, $\omega$ is contained in exactly five sets, then we have six possible, and distinct, decompositions, each corresponding to the single set which does not contain $\omega$.  These are given as follows.

First, suppose $\omega \not\in R_3$ (i.e., $\omega \in R_1 \cap R_2 \cap C_1 \cap C_2 \cap C_3$).  We cannot assume that $V_{33}(\omega) = V_{31}(\omega) V_{32}(\omega)$, but, since $\omega \in C_3$, we know that $V_{33}(\omega) = -V_{13}(\omega) V_{23}(\omega)$.  Furthermore, since $\omega$ is contained in both $R_1$ and $R_2$, we may decompose $V_{13}(\omega) = X_1(\omega) X_2(\omega)$ and $V_{23}(\omega) = Y_1(\omega) Y_2(\omega)$.  From this we conclude that
\begin{equation}
\omega \not\in R_3 \;\Rightarrow\; V_{33}(\omega) = -X_1(\omega) X_2(\omega) Y_1(\omega) Y_2(\omega).
\label{eqn:V33-}
\end{equation}
This provides a full decomposition of all nine random variables in terms of the four basic ones.  Note that the above decomposition of $V_{33}(\omega)$ will be valid whenever $\omega \in C_3 \cap R_1 \cap R_2$.

Next, suppose that $\omega \not\in C_3$ (but, again, is contained in the other five).  Now $V_{33}(\omega)$ is decomposed as follows.
\begin{equation}
\omega \not\in C_3 \;\Rightarrow\; V_{33}(\omega) = X_1(\omega) Y_2(\omega) X_2(\omega) Y_1(\omega).
\label{eqn:V33+}
\end{equation}
Of course, the order of the four factors in unimportant.  The above decomposition of $V_{33}(\omega)$ will be valid whenever $\omega \in R_3 \cap C_1 \cap C_2$.

If $\omega$ is supposed to be in all sets but $R_1$, then we can no longer decompose $V_{13}(\omega)$ as $X_1(\omega) X_2(\omega)$.  Since $\omega \in R_3 \cap C_3$, however, we may deduce that
\begin{equation}
V_{31}(\omega) V_{32}(\omega) = -V_{13}(\omega) V_{23}(\omega),
\end{equation}
and from this we conclude that
\begin{equation}
\omega \not\in R_1 \;\Rightarrow\; V_{13}(\omega) = -X_1(\omega) X_2(\omega).
\end{equation}
Note that $V_{33}(\omega)$ is decomposed according to Eqn.\ (\ref{eqn:V33+}).

Proceeding in a similar manner, find
\begin{align}
\omega \not\in R_2 \;\Rightarrow\; V_{23}(\omega) &= -Y_1(\omega) Y_2(\omega) \\
\omega \not\in C_1 \;\Rightarrow\; V_{31}(\omega) &= -X_1(\omega) Y_2(\omega) \\
\omega \not\in C_2 \;\Rightarrow\; V_{32}(\omega) &= -X_2(\omega) Y_1(\omega) \; .
\end{align}

We conclude that it may be possible to decompose any $V_{ij}(\omega)$ in terms of one or more of $X_1(\omega)$, $Y_1(\omega)$, $X_2(\omega)$, $Y_2(\omega)$, provided that $\omega$ is contained in all but one of the six row/column sets.  The decomposition is not unique, however, as it depends upon which of the six row/column sets $\omega$ is not contained in.  As discussed previously, this, in turn, will be determined by which operators one chooses to measure.
 
%------------------------------------------------------------------------------

\subsection{Huang Single-Photon Experiment}

An early experiment to test noncontextuality was performed by Huang \textit{et al.}\ \cite{Huang2003} using photon path and polarization measurements.  The concept of this experiment was based on the theoretical work of Simon \textit{et al.}\ \cite{Simon2000}, who suggested a possible test of noncontextual hidden variable theories using two degrees of freedom (path and spin) for a single spin-1/2 particle.  By measuring polarization in place of spin, the experimenters were able to perform an equivalent test using a single photon.

In the experiment, the photon is initially prepared in the entangled state
\begin{equation}
\ket{\psi} = \frac{1}{\sqrt{2}} \bigl[ \ket{u}\otimes\ket{z+} \;+\; \ket{d}\otimes\ket{z-} \bigr],
\label{eqn:Bell-Huang}
\end{equation}
where the first component corresponds to the path ($u$ = up, $d$ = down) and the second component corresponds to the polarization ($z+$ = vertical, $z-$ = horizontal).  The former correspond to eigenstates of $\hat{\sigma}_z \otimes \hat{1}$, while the latter correspond to eigenstates of $\hat{1} \otimes \hat{\sigma}_z$.  Specifically,
\begin{subequations}
\begin{align}
\hat{\sigma}_z \otimes \hat{1} &= \Bigl( \ket{u}\bra{u} - \ket{d}\bra{d} \Bigr) \otimes \hat{1} \\
\hat{1} \otimes \hat{\sigma}_z &= \hat{1} \otimes \Bigl( \ket{z+}\bra{z+} - \ket{z-}\bra{z-} \Bigr) \; .
\end{align}
\end{subequations}

Following \cite{Huang2003}, these two operators will be denoted $\hat{Z}_1$ and $\hat{Z}_2$, respectively.  In addition, the authors consider the operators $\hat{X}_1 = \hat{\sigma}_x \otimes \hat{1}$ and $\hat{X}_2 = \hat{1} \otimes \hat{\sigma}_x$.  For the particular quantum state, $\ket{\psi}$, chosen by the experimenters, quantum mechanics predicts that a measurement of either $\hat{Z}_1 \hat{Z}_2 = \hat{\sigma}_z \otimes \hat{\sigma}_z$ or $\hat{X}_1 \hat{X_2} = \hat{\sigma}_x \otimes \hat{\sigma}_x$ always results in the value $+1$.

Based on the theoretical work of Simon \textit{et al.}\ \cite{Simon2000}, the authors assert that, for systems prepared in this way, a noncontextual hidden variable theory would predict that any joint measurement of the commuting observables $\hat{Z}_1 \hat{X}_2 = \hat{\sigma}_z \otimes \hat{\sigma}_x$ and $\hat{X}_1 \hat{Z}_2 = \hat{\sigma}_x \otimes \hat{\sigma}_z$ must result in the same outcome for both observables.  Quantum mechanics predicts that the outcomes are always different.  The experimental task was to make such a measurement and ascertain whether the outcomes are indeed equal.  The result was that only about 19\% of the measurements showed identical outcomes for the two observables, in agreement with quantum mechanics and at variance with their prediction for a noncontextual theory.

The theoretical argument of Simon \textit{et al.}\ is straightforward but relies on an assumption of operator decomposability.  As is common in discussions of noncontextuality, they associate with each operator $\hat{A}$ a predetermined value $v[\hat{A}]$.  Thus, for example, $v[\hat{X}_1] = X_1(\omega)$ for some particular $\omega \in \Omega$.  The interpretation of $v[\hat{X}_1]$ is, however, subtly different from that of $X_1(\omega)$, as the former is taken to be a preexisting value which remains unchanged by the process of measurement.  By contrast, and in accordance with the interpretation of Sec.\ \ref{ssec:UC}, $X_1(\omega)$ is viewed here as the post-measurement outcome.  The difference in the two interpretations lies in whether $\omega$ is viewed as the pre- or post-measurement hidden variable state.  It is only in the former interpretation that a contradiction with quantum mechanics arises.

With this notation in mind, Simon \textit{et al.}\ observe that, for the particular choice of $\ket{\psi}$ in Eqn.\ (\ref{eqn:Bell-Huang}), the outcomes $v[\hat{Z}_1\hat{Z}_2] = +1$ and $v[\hat{X}_1\hat{X}_2] = +1$ always occur.  They then make the following decomposability assumptions:
\begin{subequations}
\begin{align}
v[\hat{X}_1\hat{X}_2] &= v[\hat{X}_1] v[\hat{X}_2] \\
v[\hat{Z}_1\hat{Z}_2] &= v[\hat{Z}_1] v[\hat{Z}_2] \\
v[\hat{X}_1\hat{Z}_2] &= v[\hat{X}_1] v[\hat{Z}_2] \\
v[\hat{Z}_1\hat{X}_2] &= v[\hat{Z}_1] v[\hat{X}_2] \; ,
\end{align}
\end{subequations}
from which one readily deduces that $v[\hat{Z}_1\hat{X}_2] = v[\hat{X}_1\hat{Z}_2]$.

The problem may be mapped to the magic square of Sec.\ \ref{sec:PS} by interchanging $\hat{\sigma}_y$ and $\hat{\sigma}_z$.  We may then define six analogous row/column sets, $R_1', \ldots, C_3'$, and nine noncontextual random variables $V_{ij}'$.  We then see that the decomposition is valid only if $\omega \in R_1' \cap R_2' \cap C_1' \cap C_2'$.  Since the actual experiment measures Row 3 (i.e., $\hat{X}_1\hat{Z}_2$ and $\hat{Z}_1 \hat{X}_2$), we are guaranteed only that $\omega \in R_3'$.  Therefore, if the measurement outcomes for the two observables are not equal, we merely conclude that $\omega \not\in R_1' \cap R_2' \cap C_1' \cap C_2'$ and, so, the decomposition was invalid.  Thus, the experimental results of Huang \textit{et al.}\ do not rule out a noncontextual hidden variable interpretation.

%------------------------------------------------------------------------------

\subsection{Hasegawa Neutron Interferometry Experiment}

In a recent experiment using neutron interferometry, Hasegawa \textit{et al.}\ \cite{Hasegawa2006} claim to have obtained empirical confirmation of the Kochen-Specker result by showing violations of a certain Bell-like inequality.  The authors consider a single-particle system for which two observables are measured: the spin (in a particular direction) and the path taken in the interferometer.  In the experiment, the system is prepared in the Bell state
\begin{equation}
\ket{\psi} = \frac{1}{\sqrt{2}} \bigl[ \ket{\downarrow}\otimes\ket{I} \;-\; \ket{\uparrow}\otimes\ket{II} \bigr],
\label{eqn:Npsi}
\end{equation}
where the first component corresponds to the spin (in the $z$ direction) and the second represents the interferometer path.

In each run of the experiment, exactly one of three observables is measured, represented here by the operators $\hat{\sigma}_x\otimes\hat{\sigma}_x, \, \hat{\sigma}_y\otimes\hat{\sigma}_y, \, \hat{\sigma}_z\otimes\hat{\sigma}_z$, where%
\begin{subequations}
\begin{align}
\hat{\sigma}_z \otimes \hat{1} &= \Bigl( \ket{\uparrow}\bra{\uparrow} - \ket{\downarrow}\bra{\downarrow} \Bigr) \otimes \hat{1} \\
\hat{1} \otimes \hat{\sigma}_z &= \hat{1} \otimes \Bigl( \ket{I}\bra{I} - \ket{II}\bra{II} \Bigr) \; .
\end{align}
\end{subequations}
Note that, for this particular choice of $\ket{\psi}$, each such measurement will, theoretically, always result in an outcome of $-1$.  For the experiment, multiple independent runs were performed to get statistical averages of each of these observables.

The resulting measured averages, denoted $E_x$, $E_y$, and $E_z$, are compared against the corresponding quantum predictions.  The empirical test consists of comparing the empirical quantity
\begin{equation}
C' := 1 - E_x - E_y - E_z
\end{equation}
against the quantum prediction
\begin{equation}
C_{\rm QM} := 1 - \bra{\psi} [\hat{\sigma}_x\otimes\hat{\sigma}_x + \hat{\sigma}_y\otimes\hat{\sigma}_y + \hat{\sigma}_z\otimes\hat{\sigma}_z] \ket{\psi} = 4,
\end{equation}
and a value, $\overline{C}_{\rm NC}$, predicted for a noncontextual hidden variable theory.  In Eqn.\ (6) of reference \cite{Hasegawa2006}, the authors predict that $|\overline{C}_{\rm NC}| \le 2$ based on a set of assumptions in Eqn.\ (2) of the same reference.  The experiment yielded a measured value of $C' = 3.138 \pm 0.015$, which clearly violates their noncontextual prediction.

In fact, the noncontextual prediction is based on a particular assumption regarding the decomposability of the measured observables.  To see this, first note that the noncontextual prediction is
\begin{equation}
\overline{C}_{\rm NC} = 1 - \sum_{i=1}^{3} \int V_{i3}(\omega) d P_i(\omega) = \int C_{\rm NC}(\omega) d P_6(\omega),
\end{equation}
where
\begin{equation}
C_{\rm NC}(\omega) = 1 - V_{13}(\omega) - V_{23}(\omega) - V_{33}(\omega)
\end{equation}
and, since the marginal distributions are noncontextual, for $i = 1,2,3$
\begin{equation}
\int V_{i3}(\omega) \, d P_i(\omega) = \int V_{i3}(\omega) \, d P_6(\omega).
\end{equation}

Now, in \cite{Hasegawa2006} the authors assume the following decomposition.
\begin{multline}
% \begin{equation}
C_{\rm NC}(\omega) = 1 - X_1(\omega) X_2(\omega) - Y_1(\omega) Y_2(\omega) \\
- X_1(\omega) X_2(\omega) Y_1(\omega) Y_2(\omega).
% \end{equation}
\end{multline}
Such a decomposition holds if and only if $\omega \in R_1 \cap R_2 \cap R_3 \cap C_1 \cap C_2$ --- i.e., $\omega \not\in C_3$ and is contained in the other five sets.  One readily verifies that $C_{\rm NC}(\omega) \in \{-2,+2\}$ for every such $\omega$.  If only such values of $\omega$ are possible, the prediction $|\overline{C}_{\rm NC}| \le 2$ is obtained.  (In fact, for the particular choice of $\ket{\psi}$ used, only $C_{\rm NC}(\omega) = 2$ will be realized; hence, this decomposition implies $\overline{C}_{\rm NC} = 2$.)

This assumption regarding $\omega$ is, however, unnecessary.  Following the discussion of Sec.\ \ref{ssec:UC}, a measurement of $V_{i3}$ would entail only that $\omega \in R_i \cup C_3$.  It is certainly possible that every such $\omega$ is not contained in $C_3$, and contained in the other five sets, but this need not be so.  It may be, for example, that $\omega \in R_1 \cap R_2 \cap C_3$, in which case we have the following, alternative decomposition. 
\begin{multline}
% \begin{equation}
C_{\rm NC}(\omega) = 1 - X_1(\omega) X_2(\omega) - Y_1(\omega) Y_2(\omega) \\
+ X_1(\omega) X_2(\omega) Y_1(\omega) Y_2(\omega).
% \end{equation}
\end{multline}
In this case, we find that $C_{\rm NC}(\omega) = \{0,4\}$, with $4$ the only possible value given the choice of $\ket{\psi}$ used in the experiment.  If the measurement process results only in such values of $\omega$, then the noncontextual prediction agrees precisely with that of quantum mechanics.  Indeed, the Sequential Measurements model of Sec.\ \ref{sec:IM} produces exactly this effect.  Thus, the experimental results of Hasegawa \textit{et al.}\ do not rule out a noncontextual hidden variable interpretation.

%------------------------------------------------------------------------------

\subsection{Proposed Experiment of Cabello \textit{et al.}}

In a related and more recent article, Cabello \textit{et al.}\ \cite{Cabello2008} suggest an alternative method of testing quantum contextuality, again, using single-neutron interferometry.  Using an experimental setup similar to that described in \cite{Hasegawa2006} and the same initial entangled state as Eqn.\ (\ref{eqn:Npsi}), they propose to perform a series five separate measurements of the following sets of observables: (1) $\hat{X}_1$, $\hat{X}_2$, (2) $\hat{Y}_1$, $\hat{Y}_2$, (3) $\hat{V}_{31}$, $\hat{X}_1$, $\hat{Y}_2$, (4) $\hat{V}_{32}$, $\hat{Y}_1$, $\hat{X}_2$, and, finally, (5) $\hat{V}_{31}$, $\hat{V}_{32}$.  In each of the five experiments, the product of the observations is taken, and the results are averaged over multiple runs.  Quantum mechanics predicts the following:
\begin{subequations}
\begin{align}
\langle \psi | \hat{X}_1 \hat{X}_2 | \psi \rangle = \langle \psi | \hat{V}_{13} | \psi \rangle &= -1 \\
\langle \psi | \hat{Y}_1 \hat{Y}_2 | \psi \rangle = \langle \psi | \hat{V}_{23} | \psi \rangle &= -1 \\
\langle \psi | \hat{V}_{31} \hat{X}_1 \hat{Y}_2 | \psi \rangle &= 1 \\
\langle \psi | \hat{V}_{32} \hat{Y}_1 \hat{X}_2 | \psi \rangle &= 1 \\
\langle \psi | \hat{V}_{31} \hat{V}_{32} | \psi \rangle = \langle \psi | \hat{V}_{33} | \psi \rangle &= -1.
\end{align}
\end{subequations}

In fact, quantum mechanics predicts that these results hold, not only on average, but for each individual (and ideal) measurement.  Based on this observation, the authors assert that a noncontextual hidden variable theory should satisfy the following relations:
\begin{subequations}
\begin{align}
X_1(\omega) X_2(\omega) &= -1 \label{eqn:3a} \\
Y_1(\omega) Y_2(\omega) &= -1 \label{eqn:3b} \\
V_{31}(\omega) X_1(\omega) Y_2(\omega) &= 1 \label{eqn:3c} \\
V_{32}(\omega) Y_1(\omega) X_2(\omega) &= 1 \label{eqn:3d} \\
V_{31}(\omega) V_{32}(\omega) &= -1 \label{eqn:3e},
\end{align}
\end{subequations}
where the authors assume (implicitly) that these relations hold for all $\omega \in \Omega$.  (See Eqns.\ (3a)--(3e) in \cite{Cabello2008}.)  They then note that no single $\omega$ can possibly satisfy all five relations, since the product of the left-hand side is $+1$, while the product of the right-hand side is $-1$.

To understand this better, let us define the sets $B_i := \{ \omega \in \Omega : V_{i1}(\omega) V_{i2}(\omega) = -1 \}$ for $i=1,2,3$.  By definition, Eqn.\ (\ref{eqn:3a}) is satisfied iff $\omega \in B_1$, Eqn.\ (\ref{eqn:3b}) is satisfied iff $\omega \in B_2$, and Eqn.\ (\ref{eqn:3e}) is satisfied iff $\omega \in B_3$.  Furthermore, Eqns.\ (\ref{eqn:3c}) and (\ref{eqn:3d}) are satisfied iff $\omega \in C_1$ and $\omega \in C_2$, respectively.  The impossibility of satisfying all five equations simultaneously implies that
\begin{equation}
B_1 \cap B_2 \cap B_3 \cap C_1 \cap C_2 = \varnothing.
\end{equation}
This result is simular to that for the six row/column sets, which were found to have no common intersection point.  It is the probabilities, however, that make this situation appear paradoxical.

For any quantum state, $P_4[C_1] = P_5[C_2] = 1$.  Furthermore, for the particular form of $|\psi\rangle$ chosen, $P_i[B_i] = 1$ for $i = 1,2,3$.  As has been argued previously, this does not, however, imply that any of these sets is identical to $\Omega$.  It is for this reason that the inequalities expressed in Eqns.\ (4) and (5) of Ref.\ \cite{Cabello2008} are invalid.  Now, it is also the case that, quite generally, $P_6[C_3] = 1$ and $P_i[R_i] = 1$ for $i = 1,2,3$.  Thus, $P_i[R_i \cap B_i] = 1$ and $P_6[C_3 \cap B_i] = 1$.  In other words, a measurement of, say, Row 1 will result in a post-measurement microstate, $\omega$, such that $X_1(\omega) X_2(\omega) = V_{13}(\omega) = -1$, while a measurement of Column 3 will result in a (possibly different) post-measurement microstate, $\omega'$, such that $V_{13}(\omega') = -1$, $V_{23}(\omega') = -1$, and $V_{33}(\omega') = -1$.  The mere fact that, say, $V_{13}(\omega) = V_{13}(\omega')$ does not imply, for example, that $X_1(\omega) = X_1(\omega')$ or $X_2(\omega) = X_2(\omega')$.

Since a noncontextual hidden variable theory does not predict that all five equations are ever satisfied, a violation of the proposed inequalities will not rule out the possibility of a noncontextual hidden variable interpretation.  Indeed, the Sequential Measurements model presented here will exactly reproduce the predicted quantum results.  

%%%%%%%%%%%%%%%%%%%%%%%%%%%%%%%%%%%%%%%%%%%%%%%%%%%%%%%%%%%%%%%%%%%%%%%%%%%%%%%

\section{Summary and Conclusions}
\label{sec:SC}

In this paper, the question of quantum contextuality in the Mermin-Peres square has been considered.  It was shown that a deterministic, noncontextual description is possible if one allows for the possibility that the hidden variable states may be disturbed through the process of measurement.  Thus, the (inaccessible) pre-measurement value of an observable may be different from its post-measurement outcome.  The Kochen-Specker theorem applies when one assumes that these two quantities are identical.

This assumption was found to follow from the Functional Composition Principle, which does not itself follow from any quantum theoretic principle.  Rather, quantum mechanics demands only that the set of hidden variable states over which a given functional relation among commuting operators holds must have a probability of 1 with respect to a distribution corresponding to the particular set of commuting operators.  This alone merely shifts the question of contextuality from the random variables to the probability measures.  One way to understand how such an apparent contextual dependency may arise is to suppose that the hidden variable states are modified through interaction between the measuring device and the system under interrogation.

That such an interpretation is possible was shown through the construction of an explicit, albeit contrived, noncontextual hidden variable model in Sec.\ \ref{sec:IM}.  One version of the model, for sequential measurements, was found to be sufficient in modeling measurements of operators from the Mermin-Peres array that are time-like separated or light-like separated and not co-located.  From this, a second version, for simultaneous measurements, was derived which models space-like separated measurements or ones that are light-like separated and co-located.  The latter is explicitly nonlocal, unless the measurements are co-located, as would be the case for single-particle systems.  Both treat measurements na\"ively as point-like space-time events.  It is an open question whether a completely local model of the Mermin-Peres square can be constructed.  A general local hidden variable theory for a four-dimensional Hilbert space appears to be ruled out by Bell's inequality.

Finally, empirical tests of quantum contextuality in two recent experiments, Huang \textit{et al.}\ \cite{Huang2003} and Hasegawa \textit{et al.}\ \cite{Hasegawa2006}, and one proposed experiment by Cabello \textit{et al.}\ \cite{Cabello2008} were considered.  In all cases, it was found that the authors' predictions for a noncontextual theory were based on an assumption of the functional composition principle and the resulting operator decomposition.  Without this assumption, none of the performed or proposed experiments are capable of ruling out a noncontextual hidden variable interpretation.  As all experiments use measurements of path and polarization/spin on a single photon/neutron, none are capable of ruling out a local hidden variable interpretation either.

%%%%%%%%%%%%%%%%%%%%%%%%%%%%%%%%%%%%%%%%%%%%%%%%%%%%%%%%%%%%%%%%%%%%%%%%%%%%%%%

\section*{Acknowledgments}

I would like to thank Drs.\ T. Yudichak, J. Gelb, and T. Lupher for many enjoyable discussions and helpful suggestions.  I would also like to acknowledge ARL:UT for its financial support under Internal Research and Development Grant No.\ 986.

%%%%%%%%%%%%%%%%%%%%%%%%%%%%%%%%%%%%%%%%%%%%%%%%%%%%%%%%%%%%%%%%%%%%%%%%%%%%%%%

\appendix

\section*{Proofs for the Sequential Measurements Model}

In what follows we shall use the following notation.  Let $\hat{\Pi}_{ij}^{\pm}$ denote the projection operator for $\hat{V}_{ij}$ onto the subspace corresponding to the eigenvalue $\pm1$.  Thus,
\begin{equation}
\hat{V}_{ij} = \hat{\Pi}_{ij}^{+} - \hat{\Pi}_{ij}^{-}.
\end{equation}

\subsection*{Proof of Theorem \ref{thm:Ordering}}

The theorem may be restated as follows.  Let $\hat{V}_{i_1 j_1}, \ldots, \hat{V}_{i_n j_n}$ be a particular sequence of measurements and let $s_1, \ldots, s_n$ be a particular sequence of corresponding outcomes.  We must show that the joint probability of this sequence according to the model, namely,
\begin{multline}
% \begin{equation}
p(s_1, \ldots, s_n) := \mu[g\circ\varphi_{i_1 j_1} = s_1, \; \ldots, \\
g\circ\varphi_{i_n j_n}\circ\cdots\circ\varphi_{i_1 j_1} = s_n],
% \end{equation}
\end{multline}
is equal to the quantum mechanical prediction, namely,
\begin{equation}
q(s_1, \ldots, s_n) := \mathrm{Tr}\left[ \hat{\rho} \; \hat{\Pi}_{i_1 j_1}^{s_1} \!\cdots\, \hat{\Pi}_{i_n j_n}^{s_n} \right].
\end{equation}

\begin{proof}
The proof is by induction.  Let $U$ be a random variable with distribution $\mu$ and let $U' := \varphi_{i_1 j_1}(U)$.  Define
\begin{equation}
K(u) := \begin{cases}
\max\{ k \ge 0 : u \in M_k \} & u \in \bigcup_{k \ge 0} M_k \\
\infty & \text{otherwise} \; .
\end{cases}
\end{equation}
For the $n = 1$ case we observe that
\begin{equation*}
\begin{split}
p(-1) &= \mu[ g(U') = -1 ] \\
&= \sum_{k=1}^{\infty} \mu[ U'_k < \tfrac{1}{2} \,|\, K(U') = k ] \; \mu[K(U') = k] \\
&= \mu[ U'_1 < \tfrac{1}{2} \,|\, K(U') = 1 ] \\
&= \mu[ U'_1 < \tfrac{1}{2} \,|\, U \in M_0 ],
\end{split}
\end{equation*}
since $K(U') = 1$ iff $U \in M_0$ and $\mu[M_0] = 1$.  Furthermore,
\begin{equation*}
\begin{split}
p(-1) &= \mu[ U'_1 < \tfrac{1}{2} \;|\; U \in M_0] \\
&= \mu[ \nu_{i_1 j_1}(\chi_{i_1 j_1}(U)) < \tfrac{1}{2} \;|\; U \in M_0 ] \\
&= \mu[ \chi_{i_1 j_1}(U) = -1 \;|\; U \in M_0] \\
&= \mu[ U_1 \le F_{i_1 j_1}(-1) \;|\; U \in M_0 ] \\
&= \mu[ U_1 \le F_{i_1 j_1}(-1) ] \\
&= F_{i_1 j_1}(-1) \\
&= \mathrm{Tr}[\hat{\rho} \; \hat{\Pi}_{i_1 j_1}^{-}].
\end{split}
\end{equation*}
Thus, $p(s_1) = q(s_1)$.  For $n > 1$, suppose
\begin{equation*}
p(s_1, \ldots, s_{n-1}) = q(s_1, \ldots, s_{n-1}).
\end{equation*}
Since
\begin{equation*}
\begin{split}
p(s_1, \dots, s_n) &= p(s_n|s_1, \ldots, s_{n-1}) p(s_1, \ldots, s_{n-1}) \\
&= p(s_n|s_1, \ldots, s_{n-1}) q(s_1, \ldots, s_{n-1}),
\end{split}
\end{equation*}
we need only show that
\begin{equation*}
p(-1|s_1, \ldots, s_{n-1}) = q(-1|s_1, \ldots, s_{n-1}).
\end{equation*}

Now, letting $V := \varphi_{i_{n-1} j_{n-1}}( \cdots \varphi_{i_1 j_1}(U) \cdots )$, $V' := \varphi_{i_n j_n}(V)$, and
\begin{equation*}
S := \bigcup_{k = 1}^{n-1} (g \circ \varphi_{i_{k} j_{k}} \circ \cdots \circ \varphi_{i_1 j_1})^{-1}[\{s_{k}\}],
\end{equation*}
we find
\begin{equation*}
\begin{split}
p(-1 &|s_1, \ldots, s_{n-1}) = \mu[ g(V') = -1 | S ] \\
&= \sum_{k=1}^{\infty} \mu[ V'_k < \tfrac{1}{2} \,|\, K(V') = k, \, S] \mu[K(V') = k \,|\, S] \\
&= \mu[ V'_n < \tfrac{1}{2} \,|\, K(V') = n, \, S] \\
&= \mu[ V'_n < \tfrac{1}{2} \,|\, V \in M_{n-1}, \, S],
\end{split}
\end{equation*}
since $K(V') = n$ iff $V \in M_{n-1}$ and $\mu[V \in M_{n-1} | S] = 1$.  Furthermore,
\begin{equation*}
\begin{split}
p(-1 &|s_1, \ldots, s_{n-1}) = \mu[ V'_n < \tfrac{1}{2} \,|\, V \in M_{n-1}, \, S] \\
&= \mu[ V_n \le F_{i_n j_n}(-1|V_1, \ldots, V_{n-1}) \,|\, S] \\
&= q(-1 | s_1, \ldots, s_{n-1}) \; .
\end{split}
\end{equation*}

\end{proof}

\subsection*{Proof of Corollary \ref{cor:Persistence}}

We must show that, for $1 \le k \le n-1$, if $\hat{V}_{i_n j_n} = \hat{V}_{i_k j_k}$ and $\hat{V}_{i_k j_k}, \ldots, \hat{V}_{i_n j_n}$ are mutually commuting, then
\begin{equation*}
p(s_n|s_1,\ldots,s_k,\ldots,s_{n-1}) = \delta(s_k,s_n).
\end{equation*}

\begin{proof}
By Theorem \ref{thm:Ordering}, we have
\begin{equation*}
\begin{split}
p(s_n&|s_1,\ldots,s_{n-1}) = q(s_n|s_1,\ldots,s_{n-1}) \\
&= \frac{\mathrm{Tr}\left[ \hat{\rho} \; \hat{\Pi}_{i_1 j_1}^{s_1} \!\cdots\, \hat{\Pi}_{i_k j_k}^{s_k} \!\cdots\, \hat{\Pi}_{i_n j_n}^{s_n} \right]}{ \mathrm{Tr}\left[ \hat{\rho} \; \hat{\Pi}_{i_1 j_1}^{s_1} \!\cdots\, \hat{\Pi}_{i_{n-1} j_{n-1}}^{s_{n-1}} \right]} \\
&= \frac{\mathrm{Tr}\left[ \hat{\rho} \; \hat{\Pi}_{i_1 j_1}^{s_1} \!\cdots\, \hat{\Pi}_{i_k j_k}^{s_k} \hat{\Pi}_{i_n j_n}^{s_n} \hat{\Pi}_{i_{k+1} j_{k+1}}^{s_{k+1}} \!\cdots\, \hat{\Pi}_{i_{n-1} j_{n-1}}^{s_{n-1}} \right]}{ \mathrm{Tr}\left[ \hat{\rho} \; \hat{\Pi}_{i_1 j_1}^{s_1} \!\cdots\, \hat{\Pi}_{i_{n-1} j_{n-1}}^{s_{n-1}} \right]} \\
&= \delta(s_k, s_n),
\end{split}
\end{equation*}
since $\hat{\Pi}_{i_k j_k}^{s_k} \hat{\Pi}_{i_n j_n}^{s_n} = \hat{\Pi}_{i_k j_k}^{s_k} \hat{\Pi}_{i_k j_k}^{s_n} = \delta(s_k,s_n) \; \hat{\Pi}_{i_k j_k}^{s_k}$.
\end{proof}

\rule[2in]{0pt}{0pt}

%%%%%%%%%%%%%%%%%%%%%%%%%%%%%%%%%%%%%%%%%%%%%%%%%%%%%%%%%%%%%%%%%%%%%%%%%%%%%%%

%%%%%%%%%%%%%%%%%%%%%%%%%%%%%%%%%%%%%%%%%%%%%%%%%%%%%%%%%%%%%%%%%%%%%%%%%%%%%%%

\end{document}